# Coherent anti-Stokes Raman Scattering Lidar Using Slow Light: A Theoretical Study

Li Gong, Haifeng Wang

Department of Physics, National University of Singapore 117542

**Abstract:**

We theoretically investigate a scheme in which backward coherent anti-Stokes Raman scattering (CARS) is significantly enhanced by using slow light. Specifically, we reduce the group velocity of the Stokes excitation pulse by introducing a coupling laser that causes electromagnetically induced transparency (EIT). When the Stokes pulse has a spatial length shorter than the CARS wavelength, the backward CARS emission is significantly enhanced. We also investigated the possibility of applying this scheme as a CARS lidar with $O_2$ or $N_2$ as the EIT medium. We found that if nanosecond laser with large pulse energy (>1 J) and a telescope with large aperture (~10 m) are equipped in the lidar system, a CARS lidar could become much more sensitive than a spontaneous Raman lidar.

## I. INTRODUCTION

Raman lidar is a powerful tool in atmospheric detection by measuring the backward spontaneous Raman scattering of atmosphere [1,2]. It can provide automated continuous measurements of a broad range of atmospheric targets including water vapor [2-4], cloud liquid water [5], atmospheric density and temperature [6,7], aerosol backscattering and extinction [3,8], ozone [1], and pollutants [1] in both troposphere and stratosphere [1,2].

However, Raman lidar suffers from low signal-to-noise ratio because of the low cross section of spontaneous Raman scattering [2]. A possible way to improve the signal level is to use nonlinear Raman scattering processes such as coherent anti-Stokes Raman scattering (CARS) and stimulated Raman scattering (SRS) [9-12]. Since the signal in these processes scales up nonlinearly with excitation intensity, they could significantly enhance the sensitivity of Raman lidar with proper laser power.

In two-color CARS and SRS, the sample is illuminated by two pulsed laser beams at different frequencies called the pump beam ($\omega_p$) and the Stokes beam ($\omega_s<\omega_p$). Raman resonance happens when $\omega_p$-$\omega_s$ matches the frequency of a molecular vibration mode. CARS detects the output field at a new frequency $\omega_a=2\omega_p$-$\omega_s$ which is significantly enhanced at Raman resonance, and SRS detects the loss of the pump beam or the gain of the Stokes beam caused by the resonant excitation of molecules to the vibrational level.

In a microscopic view, the pump pulse and the Stokes pulse stimulate every active molecule to vibrate, and the vibration leads to the emission of new fields (at $\omega_a$ in the case of CARS and at $\omega_p$ or $\omega_s$ in the case of SRS). For a single molecule, the emission is a dipole radiation that goes in forward and backward directions symmetrically. The total SRS or CARS signal is the coherent summation of emissions from all active molecules. In a homogeneous sample under copropagating beam geometry, for both SRS and CARS, the forward emissions constructively interfere, while the backward emissions destructively interfere. Therefore in total no signal is observed in the backward direction. This is different from spontaneous Raman scattering, in which the emissions from different molecules are incoherent and the total signal maintains the pattern of dipole radiation [11]. Since backward signal detection is desired in lidar applications, spontaneous Raman scattering is used in conventional Raman lidar [1-8]. SRS lidar was only proposed for counter-propagating beam geometry [13] or aerosol detection [14], because backward SRS signal can become significant when the size of the scattering particle is comparable to the wavelength of light.

In this study, we propose a scheme to enhance the backward nonlinear Raman signal by using special Stokes pulses with very slow group velocity, i.e., slow light. It can be produced by electromagnetically induced transparency (EIT) when a third coupling field is introduced [15-17]. We find that if the Stokes pulse is slow enough that its spatial length becomes smaller than the emission wavelength, the destructive interference of backward signal will largely disappear, and the backward signal will have the same order of magnitude as the forward signal. This could make nonlinear Raman lidar feasible.

Another important advantage of nonlinear Raman lidar comes from the low divergence of signal emission. In a spontaneous Raman lidar, the collected signal is inversely proportional to the square of the distance, i.e., $\sim 1/R^2$. In a CARS or SRS lidar, the divergent angle of backward emission is as small as that of the incident laser beams due to the coherent superposition of emission fields within a large beam cross section. Thus most of the backward CARS or SRS signal could be collected.

Theoretically both SRS and CARS signals are related to the third order susceptibility $\chi^{(3)}$ [11], which can be divided into a resonant part and a non-resonant part. In homogeneous medium, the forward SRS signal coherently mixed with the excitation field is free of non-resonant background due to heterodyne detection. However, in the backward direction the SRS signal is only mixed with incoherent Rayleigh scattering of the excitation laser, and so it would have the same non-resonant background as CARS signal. Moreover, Rayleigh scattering would make SRS signal more difficult to detect than CARS signal whose frequency is away from those excitation lasers. Therefore in this work, we theoretically investigate a CARS lidar scheme since it appears to be more advantageous over an SRS lidar.

## II. THEORETICAL FRAMEWORK

### A. General description of the model

As shown in Fig. 1(a), in our scheme, a pump, a Stokes and a coupling beam are sent into the sample, which is gas phase with different components mixed homogeneously. Firstly, we choose one of the gas components as the EIT medium. The frequency of the Stokes light should match the energy gap between the ground state |0> and an excited state |1> of the EIT medium, while the frequency of the coupling light should match the energy gap between |1> and a coupling state |2> [Fig. 1(b)]. |2> can be either higher or lower than state |1>. Based on the principle of EIT, the coupling light at a proper intensity can significantly decrease the group velocity of the Stokes light [18,19]. Secondly, we choose the wavelength of the pump light based on the characteristic Raman band of the target molecules to be detected, which we call the Raman medium. $\omega_p$-$\omega_s$ should match the frequency of the Raman band [Fig. 1(c)]. The Raman medium is assumed to be different from the EIT medium to avoid complications. For example, in Earth's atmosphere, we can choose oxygen as the EIT medium and water vapor or carbon-dioxide etc. as the Raman medium. The parameters of the lasers and the sample will be discussed in detail later.

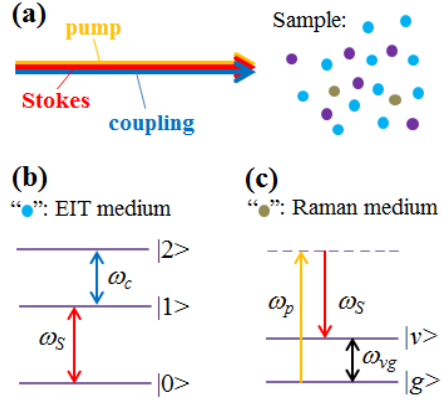

FIG. 1. (a) Scheme of CARS lidar with a normal pump pulse and a slow Stokes pulse whose group velocity is controlled by a coupling beam. The different gas components are labeled by different colors. (b) Energy diagram of EIT medium. (c) Energy diagram of Raman medium.

### B. Generation of backward CARS signal

The CARS field $\mathbf{E}_a$ is determined by the following wave equation:

$$\frac{\partial^2 \mathbf{E}_a}{\partial x^2} + \frac{\partial^2 \mathbf{E}_a}{\partial y^2} + \frac{\partial^2 \mathbf{E}_a}{\partial z^2} - \frac{n_a^2}{c^2}\frac{\partial^2 \mathbf{E}_a}{\partial t^2} = \frac{1}{c^2 \varepsilon_0}\frac{\partial^2 \mathbf{P}_a^{(3)}}{\partial t^2}, \quad (1)$$

where $\mathbf{P}_a^{(3)}$ is the third order nonlinear polarization at frequency $\omega_a=2\omega_p-\omega_S$; $c$ is the speed of light in vacuum; $\varepsilon_0$ is the vacuum permittivity; $n_a$ is the refractive index of the CARS field in the sample. Similarly, we use $n_p$, $n_S$ and $n_c$ to denote the refractive indices of the pump, Stokes and coupling fields, respectively.

We assume that both the pump and the Stokes pulses have Gaussian intensity profile propagating in $+z$ direction. We also assume that the beam diameters are much larger than the laser wavelengths and so the beams have very small divergence, which is always the case in atmospheric lidar system [2]. Thus the laser beams and the CARS signal can be approximated as plane waves. This allows us to drop the partial derivatives with respect to $x$ and $y$ in Eq. (1).

The refractive index of air at standard condition is around 1.0003, and the difference between group velocity and phase velocity of light is around $1\times 10^{-5}$ [20]. Thus we can consider the group velocity of the pump pulse to be equal to its phase velocity $c/n_p$. The group velocity of the Stokes pulse is denoted as $v_{g,S}$ which is much slower. Therefore, the pump pulse is always catching up with the Stokes pulse when they interact in the sample. If we set $z=0$ and $t=0$ as the point where the center of the pump pulse meets the center of the Stokes pulse, we have:

$$P_{a,i}^{(3)} = \varepsilon_0 \sum_{j,k,l} \chi_{ijkl}^{(3)} A_{p,j} A_{p,l} A_{S,k}^* \exp\left\{-\frac{4\ln 2(t-zn_p/c)^2}{t_p^2} - \frac{2\ln 2(t-z/v_{g,S})^2}{t_S^2} - i[\omega_a t - (\Delta k + k_a)z]\right\}, \quad (2)$$

where the indices $i$, $j$, $k$, $l$ can be $x$ or $y$; $A_p$, $A_S$ and $t_p$, $t_S$ are the field amplitude and pulse duration (full width at half maximum (FWHM) of intensity) of the pump and Stokes pulses, respectively; $k_p$, $k_S$ and $k_a$ are wave vectors of pump, Stokes and CARS fields, respectively; $\Delta k = 2k_p - k_S - k_a$ is the wave vector mismatch. On the other hand, Eq. (1) can be formally solved:

$$\mathbf{E}_a(z,t) = -\frac{n_a}{2c\varepsilon_0}\int_{-\infty}^{z}\frac{\partial}{\partial t}\mathbf{P}^{(3)}\left(z', t-\frac{z-z'}{c}n_a\right)dz' - \frac{n_a}{2c\varepsilon_0}\int_{z}^{\infty}\frac{\partial}{\partial t}\mathbf{P}^{(3)}\left(z', t+\frac{z-z'}{c}n_a\right)dz'. \quad (3)$$

The first term of the right-hand side describes a forward propagating signal, while the second term describes a backward propagating signal. In order to achieve a spectral resolution that matches the bandwidth of a typical Raman band, the pulse duration of pump and Stokes pulses should be in the order of picosecond, which is much larger than the period of an optical cycle. Thus $\partial/\partial t$ can be replaced by $-i\omega_a$ in Eq. (3). Then we obtain the signal field by substituting Eq. (2) into Eq. (3). The results are:

The forward signal:

$$\mathbf{E}_a^{FW}(z,t) = \mathbf{C}z_- \exp\left\{-\left[\frac{4\ln 2}{t_p^2}+\frac{2\ln 2}{t_S^2}-\frac{1}{4t_-^2}\right](t-zn_a/c)^2 -i\left(\omega_a-\frac{\Delta k}{2t_-}z_-\right)(t-zn_a/c)-\frac{z_-^2\Delta k^2}{4}\right\} \\ \times \int_{-\infty}^{\alpha_-}\exp\left[-\left(x-\frac{i}{2}\Delta k z_-\right)^2\right]dx \quad (4)$$

The backward signal:

$$\mathbf{E}_a^{BW}(z,t) = \mathbf{C}z_+ \exp\left\{-\left[\frac{4\ln 2}{t_p^2}+\frac{2\ln 2}{t_S^2}-\frac{1}{4t_+^2}\right](t+zn_a/c)^2 -i\left(\omega_a-\frac{\Delta k+2k_a}{2t_+}z_+\right)(t+zn_a/c)-\frac{z_+^2(\Delta k+2k_a)^2}{4}\right\} \\ \times \int_{\alpha_+}^{\infty}\exp\left[-\left(x-\frac{i}{2}(\Delta k+2k_a)z_+\right)^2\right]dx \quad (5)$$

In Eq. (4) and (5),

$$C_i = \frac{i\omega_a n_a}{2c}\sum_{j,k,l}\chi_{ijlk}^{(3)}A_{p,j}A_{p,l}A_{S,k}^*, \quad (6a)$$

$$z_\pm = \left[\frac{4\ln 2}{t_p^2 c^2}(n_p \pm n_a)^2 + \frac{2\ln 2}{t_S^2}\left(\frac{1}{v_{g,S}}\pm\frac{n_a}{c}\right)^2\right]^{-1/2}, \quad (6b)$$

$$t_\pm = \left[\frac{8\ln 2}{t_p^2 c}(n_p \pm n_a) + \frac{4\ln 2}{t_S^2}\left(\frac{1}{v_{g,S}}\pm\frac{n_a}{c}\right)\right]^{-1} z_\pm^{-1}, \quad (6c)$$

$$\alpha_\pm = \left(\frac{1}{z_\pm}+\frac{n_a}{2t_\pm c}\right)z - \frac{t}{2t_\pm}, \quad (6d)$$

In order to reveal a clearer physical picture, we set the wave vector mismatch $\Delta k$ to be zero. We will show that this is a proper assumption later. The forward and backward signal becomes:

$$\mathbf{E}_a^{FW}(z,t) = z_- \mathbf{C}\exp\left[-\frac{4\ln 2}{t_p^2}(t-n_a z/c)^2 - i\omega_a(t-n_a z/c)\right]\int_{-\infty}^{\alpha_-}e^{-x^2}dx, \quad (7)$$

$$\mathbf{E}_a^{BW}(z,t) = e^{-k_a^2 z_+^2} z_+ \mathbf{C}\exp\left[-\frac{4\ln 2(c-v_{g,S})^2}{8t_S^2 v_{g,S}^2 + t_p^2(c+v_{g,S})^2}(t+n_a z/c)^2 - i\omega_a\left(1-\frac{8t_S^2+2t_p^2(1+c/v_{g,S})}{8t_S^2+t_p^2(1+c/v_{g,S})^2}\right)(t+n_a z/c)\right] \\ \times \int_{\alpha+}^{\infty}\exp\left[-(x-ik_a z_+)^2\right]dx \quad (8)$$

The forward signal always has a non-zero Gaussian profile as expected. But the backward signal is weakened by destructive interference, which is mostly reflected in the amplitude coefficient $z_+\exp[-(k_a z_+)^2]$. To maximize the backward signal, we need to maximize $z_+\exp[-(k_a z_+)^2]$, which gives $z_+ = 1/(\sqrt{2}k_a)$. If the CARS signal is in visible region, $k_a$ is in the order of 10 μm$^{-1}$. As mentioned before, $t_p$ and $t_S$ should be at least in the order of picosecond. Then Eq. (6b) requires $v_{g,S}/c$ to be less than $10^{-4}$. Since $v_{g,S} \ll c$, we can further simplify Eqs. (6b-d) to be:

$$z_\pm = \frac{t_S v_{g,S}}{\sqrt{2\ln 2}}, \tag{9a}$$

$$t_\pm = \frac{t_S}{2\sqrt{2\ln 2}}, \tag{9b}$$

$$\alpha_\pm = \frac{\sqrt{2\ln 2}}{t_S}\left(\frac{z}{v_{g,S}} - t\right). \tag{9c}$$

After the interaction is over, the forward and backward signal will propagate away from the slow Stokes pulse. In the forward region, $z \gg v_{g,S} t_S$, we have $\alpha_- \gg 1$; in the backward region, $z \ll -v_{g,S} t_S$, we have $\alpha_+ \ll -1$ and $k_a z_+ \ll |\alpha_+|$. Therefore the integral in Eqs. (7) and (8) can be regarded as from $-\infty$ to $\infty$. The forward and back ward CARS signal finally become:

$$\mathbf{E}_a^{FW}(z,t) = t_S v_{g,S}\sqrt{\frac{\pi}{2\ln 2}}\mathbf{C}\exp\left[-\frac{4\ln 2}{t_p^2}(t - n_a z/c)^2 - i\omega_a(t - z/c)\right], \tag{10}$$

$$\mathbf{E}_a^{BW}(z,t) = \exp\left[-\frac{(k_a t_S v_{g,S})^2}{2\ln 2}\right] t_S v_{g,S}\sqrt{\frac{\pi}{2\ln 2}}\mathbf{C}\exp\left[-\frac{4\ln 2}{t_p^2}(t + n_a z/c)^2 - i\omega_a\left(1 - \frac{2v_{g,S}}{c}\right)(t + z/c)\right]. \tag{11}$$

The frequency of the forward CARS signal is $\omega_a$, while the frequency of the backward CARS signal is red shifted by a factor of $1 - 2v_{g,S}/c$. It corresponds to the Doppler red shift induced by a radar object moving at a speed of $v_{g,S}$ away from the observer. The first exponential factor in Eq. (11) gives us the ratio between backward and forward CARS signal. $v_{g,S} t_S$ can be regarded as the spatial length of the Stokes pulse.

Using Eqs. (7) and (8), the space-time profile of the CARS signal is calculated and shown in Fig. 2 alongside the profiles of the pump and Stokes pulses. In Fig. 2(a), $v_{g,S}$ is set to $0.5c$. In this case, $k_a v_{g,S} t_S \gg 1$, and the backward signal is almost zero. Only forward CARS signal appears. In Fig. 2(b), $v_{g,S}$ is set to $2.75\times10^{-5}c$. In this case, $k_a v_{g,S} t_S = \sqrt{\ln 2}$, i.e., the spatial length of the Stokes pulse is only $\sqrt{\ln 2}/2\pi$ times the wavelength of the CARS signal. The backward signal becomes $1/e$ times the forward signal in terms of intensity. $n_a$ is always set to 1.

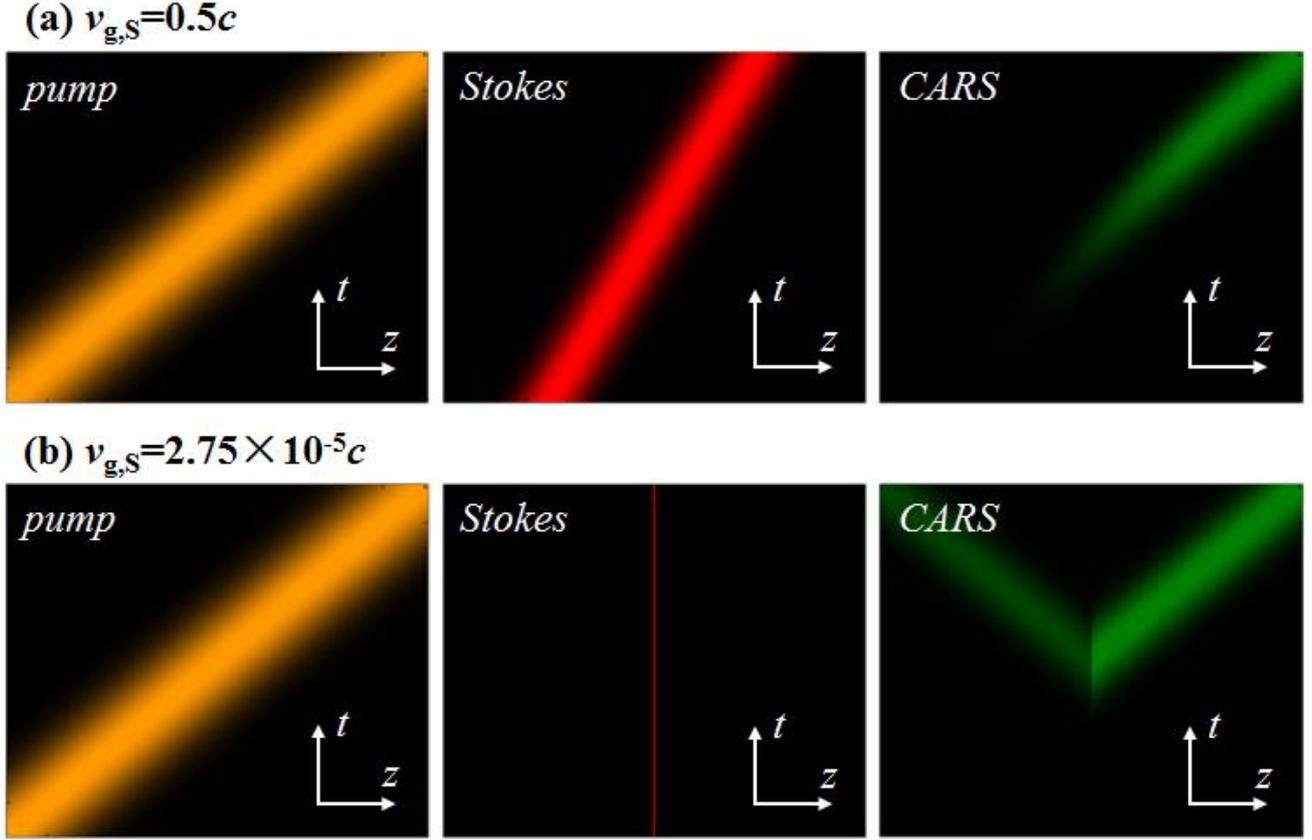

FIG. 2. The normalized intensity of pump, Stokes and CARS fields as a function of space and time. (a) $v_{g,S}=0.5c$, (b) $v_{g,S}=2.75\times10^{-5}c$. In both (a) and (b), the group velocity of the pump pulse is equal to $c$. $t_p=t_S=8$ ps. $\lambda_a=600$ μm. $z$ and $t$ in all the figures are from -4.8 mm to 4.8 mm and from -16 ps to 16 ps, respectively. Both $z$ and $t$ are set to be zero where the center of the pump pulse meets the center of the Stokes pulse. In case (a), comparing to forward CARS fields, the backward CARS signal is almost zero. In case (b), both forward and backward CARS fields have the same order of magnitude.

Fig. 3 plots the peak intensity of both forward and backward signals as a function of $v_{g,S}t_S/\lambda_a$. The backward signal achieves its maximum when:

$$v_{g,S}t_S = \frac{\sqrt{\ln 2}}{k_a} = \frac{\sqrt{\ln 2}}{2\pi}\lambda_a \approx 0.13\lambda_a. \tag{12}$$

The backward signal intensity starts to decrease quickly when $v_{g,S}t_S$ goes beyond $0.2\lambda_a$ and it is reduced to 1% of the peak value when $v_{g,S}t_S=0.31\lambda_a$. Since $v_{g,S}t_S k_a$ is in the order of 1, $v_{g,S}t_S\Delta k$ should be a very small value. Thus dropping the wave vector mismatch in Eqs. (7) and (8) is a proper assumption.

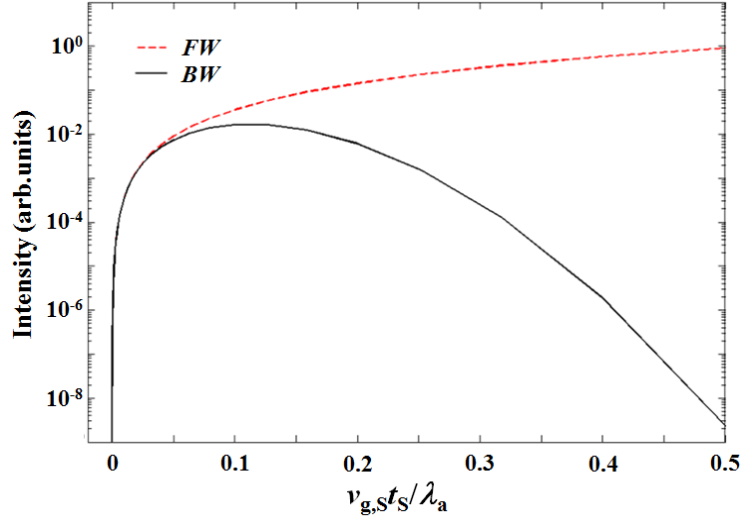

FIG. 3. The peak intensity of the both forward (FW, red dashed curve) and backward (BW, black solid curve) signals as a function of $v_{g,S}t_S/\lambda_a$.

This phenomenon can be understood intuitively this way. All the active molecules emit CARS signal in both forward and backward direction as dipole radiation. However, for an observer at backward position, the phase of the dipole emission from molecules between $z$ to $z+\lambda_a/2$ spans from 0 to $2\pi$, and thus the coherent summation of the signal largely cancels. Only when spatial length $v_{g,S}t_S$ is shorter than half wavelength, the coherent summation of the backward signal becomes relatively significant. It is exactly similar to the generation of backward CARS signal by a very short sample due to relaxed phase matching condition, which is studied both theoretically and experimentally in Ref [21].

We shall note that it is not a violation of the conservation of momentum. Although by generating a backward CARS photon, the momenta of optical fields are no longer conserved, but it is only because there is a momentum transfer between light and the medium. This is similar to the phenomena in Doppler cooling [22], in which an atom absorbs photons with a certain momentum and emits photons in random directions. The momenta of photons are not conserved, but the total momenta of the system are still conserved because of the momenta transfer from photons to the atom. In our model, we can also rigorously prove that the momenta transferred to the Raman medium are equal to the momenta loss of photons, which is presented in the Appendix. We should further emphasize that the momentum transfer between light and the medium is only significant when the interaction region is much shorter than light wavelength. If the medium is uniform and the interaction region is much longer than light wavelength, the momentum transfer at different spatial locations will cancel out, as our theory will point out at the end of the Appendix. It would lead to the strict momentum conservation (also known as the phase matching condition) of light fields, and backward emissions would vanish as seen in the right end of Fig. 3.

### C. Generating slow light by EIT

The group velocity of the Stokes light can be expressed by:

$$v_{g,S} = \frac{c}{n_S + \omega_S \frac{\partial n_S}{\partial \omega_S}}. \qquad (13)$$

$n_S$ can be calculated by:

$$n_S = \sqrt{1+\text{Re}\left[\chi^{(1)}(\omega_S)\right]}, \tag{14}$$

where $\chi^{(1)}$ is the first order susceptibility. If the Stokes light can be regarded as a perturbation to the transition from state |1> to |2> of the EIT medium as shown in Fig. 1(b), we have [11]:

$$\chi^{(1)}(\omega_S) = \frac{N_{EIT}|\mu_{23}|^2}{\hbar\varepsilon_0} \frac{\omega_S + \omega_c - \omega_{31} + i\gamma_{31}}{|\Omega_c|^2 - (\omega_S - \omega_{21} + i\gamma_{21})(\omega_S + \omega_c - \omega_{31} + i\gamma_{31})}, \tag{15}$$

where $N_{EIT}$ is the number density of the EIT molecules; $\mu_{ij}$, $\omega_{ij}$ and $\gamma_{ij}$ are the dipole momentum, transition frequency and decoherence rate between state |i> and |j>, respectively; $\Omega_c$ is the Rabi frequency of the coupling light. Under the condition of EIT:

$$|\Omega_c|^2 \gg |(\omega_S - \omega_{21} + i\gamma_{21})(\omega_S + \omega_c - \omega_{31} + i\gamma_{31})|. \tag{16}$$

Thus when $\omega_S+\omega_c$ matches $\omega_{31}$, we have:

$$\text{Re}\left[\chi^{(1)}(\omega_S)\right] = 0, \tag{17,a}$$

$$\frac{\partial}{\partial \omega_S}\text{Re}\left[\chi^{(1)}(\omega_S)\right] = \frac{N_{EIT}|\mu_{23}|^2}{\hbar\varepsilon_0|\Omega_c|^2}. \tag{17,b}$$

According to Eq. (17a), the phase velocity of the Stokes light is still $c$, i.e., the refractive index $n_S=1$. Substitute Eqs. (14) and (17) into Eq. (13), we have:

$$v_{g,S} = \frac{c}{\left(1+\frac{N_{EIT}|\mu_{23}|^2 \omega_S}{2\hbar\varepsilon_0|\Omega_c|^2}\right)} \approx \frac{2c\hbar\varepsilon_0|\Omega_c|^2}{N_{EIT}|\mu_{23}|^2 \omega_S} = \frac{4I_c}{N_{EIT}\hbar\omega_S}, \tag{18}$$

where $I_c$ is the intensity of the coupling laser. Eq. (18) shows that the group velocity can be reduced when $|\Omega_c|$ is smaller. The "≈" holds when $c \gg v_{g,S}$, and this is always the case in our study. This is the principle of how slow light is generated by EIT.

One may noticed that in our case the spatial length of the Stokes pulse $v_{g,S}t_S$ is smaller than wavelength [Eq. (12)]. Since EIT does not change both the amplitude and the phase of the spectrum, i.e. does not change the carrier frequency $\omega_S$ and the pulse duration $t_S$. The spatial length becomes smaller than wavelength because $v_{g,S}$ becomes very small.

### D. Maximizing the backward CARS signal

Now, we can examine what conditions should the EIT medium and the lasers satisfy to maximize the backward signal. First of all, according to inequality (16) and Eq. (18), there is a lower limit of $v_{g,S}$ which can be analyzed in the following way. The coupling laser is assumed to be narrowband, and thus both $\omega_S-\omega_{21}$ and $\omega_S+\omega_c-\omega_{31}$ are at the same order as the bandwidth of the Stokes laser $\Delta\omega_S$, which can be expressed by $\Delta\omega_S=2\sqrt{2}\ln2/t_S$. The decoherence of the gas phase medium is dominated by the collision of molecules when its pressure is larger than $10^{-3}$ atm [23], which has nothing to do with the energy levels of the molecules. Thus we assume $\gamma_{21}=\gamma_{31}=\gamma$. Then, we consider inequality (16) in two cases. Case (i): if $\Delta\omega_S \gg \gamma$, the lower limit of the Rabi frequency of the coupling field $|\Omega_c|_{\min} \propto \Delta\omega_S$. The lower limit of the group velocity $v_{g,S,\min} \propto (|\Omega_c|_{\min})^2 \propto \Delta\omega_S^2 \propto (1/t_S)^2$. Thus we have

$v_{g,S,min} \times t_S \propto 1/t_S$. In this case, a longer pulse duration $t_S$ or smaller $\Delta\omega_S$ is better to reduce $v_{g,S}t_S$. Case (ii): when $\Delta\omega_S$ becomes comparable with or even smaller than $\gamma$, we roughly have $|\Omega_c|_{min} \propto \gamma$. Then, $v_{g,S,min} \propto (|\Omega_c|_{min})^2 \propto \gamma^2$. Thus we have $v_{g,S,min} \times t_S \propto t_S$. In this case, a shorter pulse duration $t_S$ or larger $\Delta\omega_S$ is better to reduce $v_{g,S}t_S$. Overall, the best choice is to set $\Delta\omega_S = \gamma$, i.e.:

$$t_S = 2\sqrt{2}\ln 2/\gamma. \tag{19}$$

And in practice, we consider that the following condition is enough to satisfy inequality (16).

$$|\Omega_c|^2 \geq 10\gamma^2 \tag{20}$$

Therefore, the lower limit of the spatial length of the Stokes pulse becomes:

$$v_{g,S}t_S \geq \frac{40\ln 2\sqrt{2}c\hbar\varepsilon_0\gamma}{N_{EIT}|\mu_{23}|^2\omega_S}. \tag{21}$$

According to Eq. (12), in order to maximize the backward signal, this lower limit should be smaller than or equal to $\sqrt{ln2}\lambda_a/(2\pi)$. Since the decoherence is dominated by collision, $\gamma$ can be estimated by the mean speed $v_{mean}$ and the mean free path $l_{mean}$ of the EIT molecules:

$$\gamma = \frac{v_{mean}}{l_{mean}} = \frac{\sqrt{\frac{8k_BT}{\pi m_{EIT}}}}{\left(\frac{k_BT}{\sqrt{2}\pi d_{EIT}^2 P}\right)} = 4d_{EIT}^2 P\sqrt{\frac{\pi}{m_{EIT}k_BT}}, \tag{22}$$

where $T$ is the temperature; $P$ is the pressure; $k_B$ is Boltzmann constant; $m_{EIT}$ and $d_{EIT}$ are the mass and diameter of the EIT molecule, respectively. Substituting Eq. (22) into inequality (21) and combining with Eq. (12), we obtain the requirement for the dipole moment of EIT molecules:

$$\begin{aligned}|\mu_{23}|^2 &\geq 160\sqrt{2\ln 2}\pi\frac{\hbar\varepsilon_0 d_{EIT}^2 P}{N_{EIT}\sqrt{m_{EIT}k_BT}}\frac{\lambda_S}{\lambda_a} \\ &= 160\sqrt{2\ln 2}\pi\frac{\hbar\varepsilon_0 d_{EIT}^2\sqrt{k_BT}}{\beta\sqrt{m_{EIT}}}\frac{\lambda_S}{\lambda_a}\end{aligned}. \tag{23}$$

In the last step, $P=Nk_BT$ is used. $\beta$ is the ratio between $N_{EIT}$ and total number density $N$. We'll see in the following section that Eq. (23) is a reasonable requirement.

**III. APPLICATION FOR ATMOSPHERIC CARS LIDAR**

A. Practical limits of the EIT medium and the laser intensity

Since $\chi^{(1)}$ is used in slow light generation, the EIT medium should be considered as continuous medium rather than discrete molecules, i.e., $N_{EIT}\lambda^3 \gg 1$. Assuming that the Stokes and the coupling lasers are in visible region, $N_{EIT}\lambda^3$ from sea level to 25 km in the stratosphere is in the region of $2\times10^6 \sim 5\times10^4$ ($5\times10^5 \sim 1\times10^4$) if $N_2$ ($O_2$) is chosen as the EIT medium [24]. And so either nitrogen or oxygen can act as the EIT medium. The Raman medium is not restricted by such requirement, although $\chi^{(3)}$ is used in CARS signal generation. In principle CARS signal can be generated by even a single molecule.

As a requirement of EIT, the intensity of the Stokes laser should be small enough as a perturbation of the transition from |1> to |2> in EIT medium, i.e., its Rabi frequency $|\mu_{12}E_S/\hbar|$ should be much smaller than the

decoherent rate $\gamma$. We consider that

$$|\Omega_S| = \left|\frac{\mu_{12}}{\hbar}\sqrt{\frac{2I_S}{c\varepsilon_0}}\right| = \gamma/10, \quad (24)$$

as the upper limit of the Stokes Rabi frequency. At the same time, in order to obtain stronger CARS signal, higher pump and Stokes laser intensity is desired. Thus, it is better to choose a state |2> with a small $\mu_{12}$, so that $I_S$ can be large while the Rabi frequency is still small. For example, in oxygen, there is a weak absorption peak at 688.4 nm, which is the transition from ground state $X^3\Sigma_g^-$ to state $b^1\Sigma_g^+$ [25]. Its $\mu_{12}=2.67\times10^{-6}$ e·nm from its absorption cross section [25].

In addition, if the intensity of any beam exceeds $10^{13}\sim10^{14}$ W/cm$^2$, the ionization of $O_2$ and $N_2$ will happen, which may degrade the EIT process [26]. This is an upper limit of laser intensity.

### B. Strategy of choosing parameters

Since there are a lot of parameters involved in our scheme, now we propose a strategy to determine these parameters by the following sequence: (i) choose proper states |2> and |3> to satisfy inequality (23), and at the same time the transition from |1> to |2> should be weak; (ii) determine $t_S$ by Eq. (19), $I_c$ by Eqs. (12) and (18), and the upper limit of $I_S$ by Eq. (24); (iii) $I_p$ can be as large as possible within the ionization limit; (iv) $t_p$ can be as large as possible, until $v_{g,S}t_p$ reaches the spatial resolution of the lidar (since at least $v_{g,S}/c<10^{-4}$, if the desired spatial resolution is 3 m, $t_p$ can be > 0.1 ms); (v) the coupling laser can be a CW laser or pulsed laser whose pulse duration is much longer than $t_p$, since $v_{g,S}$ needs to be constant throughout the pump-Stokes interaction.

The molecular mass and diameter are $5.32\times10^{-26}$ kg and 0.358 nm for $O_2$, and $4.65\times10^{-26}$ kg and 0.370 nm for $N_2$, respectively [27]. Based on these numbers, we calculated $\gamma$, the minimum $|\mu_{23}|$, $t_S$, $v_{g,S}/c$ and $I_c$ under different temperature and pressure conditions (Table 1). $\lambda_S/\lambda_a$ is assumed to be 1.5. $v_{g,S}/c$ is calculated by making $v_{g,S}t_S=(\sqrt{ln2}/2\pi)\lambda_a$ with $\lambda_a=500$ nm [Eq. (12)], i.e., the backward signal is maximized. The upper limit of $I_S$ strongly depends on $\mu_{12}$, and we used $\mu_{12}=2.67\times10^{-6}$ e·nm which is mentioned above as an example.

Table 1 (The conditions listed in the left column is for tropical region in summer [24])

| | $\gamma$ (ns$^{-1}$) | $|\mu_{23}|$ (e·nm) | $t_S$ (ns) | $v_{g,S}/c$ | $I_c$ (W/cm$^2$) | $I_S$ (W/cm$^2$) |
|---|---|---|---|---|---|---|
| EIT medium | $O_2/N_2$ | $O_2/N_2$ | $O_2/N_2$ | $O_2/N_2$ | $O_2/N_2$ | $O_2$ |
| T=300 K, P=1 atm (Sea level) | 6.21/7.09 | 5.57/3.09×10$^{-2}$ | 0.315/0.277 | 5.80/6.62×10$^{-7}$ | 7.14/30.3×10$^7$ | 3.11×10$^{13}$ |
| T=270 K, P=0.5 atm (5 km, troposphere) | 3.27/3.74 | 5.43/3.01×10$^{-2}$ | 0.599/0.525 | 3.06/3.49×10$^{-7}$ | 2.09/8.86×10$^7$ | 8.64×10$^{12}$ |
| T=210 K, P=0.2 atm (15 km, tropopause) | 1.48/1.69 | 5.10/2.83×10$^{-2}$ | 1.32/1.16 | 1.39/1.58×10$^{-7}$ | 4.87/20.7×10$^6$ | 1.78×10$^{12}$ |
| T=220 K, P=0.02 atm (25 km, stratosphere) | 0.145/0.166 | 5.16/2.86×10$^{-2}$ | 13.5/11.8 | 1.36/1.55×10$^{-8}$ | 4.55/19.3×10$^4$ | 1.70×10$^{10}$ |

Table 1 tells us that the required minimum dipole momentum $|\mu_{23}|$ is one order of magnitude less than the product of electron charge and molecular diameter, which is the approximate theoretical upper limit of dipole moment. Thus it seems possible to find a proper coupling state in $O_2$ or $N_2$. Detailed studies of energy levels in $O_2$ and $N_2$ are still needed to identify such a state. In EIT experiments, $v_{g,S}/c$ can be as low as $5.6\times10^{-8}$ in cold atomic gas [18], $3.0\times10^{-7}$ in hot dilute atomic gas [28], $1.5\times10^{-7}$ in Pr doped $Y_2SiO_5$ [19], $1.9\times10^{-7}$ in ruby crystal at room temperature [29]. In atmosphere, based on our calculation, if $|\mu_{23}|$ is properly chosen, $v_{g,S}/c$ can be suppressed to $10^{-7}\sim10^{-8}$.

### C. Comparing the signal level between CARS lidar and spontaneous Raman lidar

In this subsection, we compare the signal level between a CARS lidar and a spontaneous Raman lidar at different laser pulse energies. For spontaneous Raman lidar, the signal power can be calculated as a function of pulse energy $E$ by Eq. (25):

$$P_{SR} = \sigma N_R \frac{\pi D^2}{4R^2} cE, \qquad (25)$$

where $\sigma$ is the Raman scattering cross section which is typically $10^{-29}$ cm$^2$/sr/molecule, when the excitation wavelength is at 488 nm [30] and 337.1 nm [31], $D$ is the aperture diameter of the receiver telescope, $R$ is the distance of the target from lidar. The number density of the Raman medium $N_R$ can be calculated by $N_0 P_R T_0/(P_0 T_R)$, where $N_0$=Avogadro's constant/22.4 L is the number density of air at standard pressure ($P_0$=1 atm) and temperature ($T_0$=273 K). $P_R$ and $T_R$ are the partial pressure and temperature of Raman medium, respectively. $P_R$ is calculated by air pressure at the target times the percentage of the Raman medium. Assuming the altitude of the lidar system is on sea level, $R$ becomes the altitude of the target. Thus $T_R$ can be estimated by 300 K−6 K/km×$R$.

For CARS lidar, $t_S$ is set to be 0.5 ns, as indicated in Table 1. $t_p$ is set to be equal to $t_S$. $\chi^{(3)}$ of the air is $1.7\times10^{-25}$ m$^2$/V$^2$ [11]. Thus $\chi^{(3)}$ of the Raman medium can be estimated by $1.7\times10^{-25}$ m$^2$/V$^2$ times $N_R/N_0$. Since CARS process prefers higher laser intensity, we assume that the focal spot of the telescope is at the target. If the same telescope is used to focus the laser into the atmosphere, the focal spot radius can be calculated by the Airy disk radius $r=0.61\lambda R/D$, assuming that the laser beam fills the telescope aperture. We assume that the wavelengths of pump, Stokes and CARS fields are all close to 500 nm, which is the value of $\lambda$ here. Then, the power of the backward CARS signal generated at the target can be estimated by multiplying the focal spot area and the CARS intensity:

$$P_{CARS} = \pi r^2 \varepsilon c \left| E_a^{BW} \right|^2 / 2 \qquad (26)$$

where $E_a^{BW}$ can be calculated by Eq. (11). To evaluate Eq. (11) via Eq. (6a), the peak electric field amplitudes of pump and Stokes lasers are needed. They are calculated from the peak intensities of the laser beams, which are estimated by pulse energy/pulse duration/($\pi r^2$). Since the divergent angle of CARS signal should be almost the same as the convergent angle of the incident beams, we assume 100% collection efficiency. Thus Eq. (26) is considered as the received signal power.

Fig. 4 compares the signal power between spontaneous Raman lidars and CARS lidars with different parameters. For CARS lidars, the pulse energy means the total pulse energy of pump and Stokes lasers. In each figure, the results of Raman medium with different percentage are plotted. The percentages are set to be 78%, 21% and 0.2%, which are the typical values for nitrogen, oxygen and water vapor, respectively. In Fig. 4(a-c), the pulse energy is $10^{-1.5}\sim10^{0}$ J, and the diameter of the receiver telescope is $D$=1 m. These are typical values for spontaneous Raman lidar [2]. $R$ is set to be 2 km, 5 km and 10 km, respectively. At such altitudes, the air pressure drops to 0.8 atm, 0.5 atm, and 0.3 atm, respectively. In this region, the signal of CARS lidar is weaker than that of spontaneous Raman lidar.

However, the CARS signal is proportional to the third power of laser intensity, while the spontaneous Raman signal is linearly proportional to the laser power. Therefore, higher peak intensity will benefit CARS lidars more than spontaneous Raman lidars. To increase the peak intensity of laser pulses, we could use lasers with higher power or use a telescope with larger aperture (e.g. optical reflective telescope, whose aperture can be as large as 10 m) to achieve a smaller focal spot in the atmosphere. If the diameter of the aperture $D$ increases by a factor of 10, the focal spot area $\pi r^2$ will decrease by a factor of $10^2$, and the CARS intensity will increase by a factor of $10^6$, then, the CARS power will increase by a factor of $10^4$. At the same time, the spontaneous Raman signal will increase by a factor of $10^2$ merely due to a larger receiving area. As we can see from Fig. 4(d-f), if a nanosecond laser with ~10

J pulse energy and a telescope with big aperture ($D$=10 m) are adopted in a CARS lidar, it will be much more sensitive than a spontaneous Raman lidar.

On the other hand, the CARS lidar signal decreases more rapidly than the spontaneous Raman lidar signal when distance $R$ increases. This is because the Airy disk radius r is proportional to R. Following similar discussion as above, if $R$ increases by a factor of 10, the CARS power will decrease by a factor of $10^4$, while the spontaneous Raman signal will decrease by a factor of $10^2$ due to a smaller receiving solid angle. This trend can be seen in Fig. 4 as well. Still, within our model, when D=10 m the CARS signal is largely better than spontaneous Raman signal even for a distance of 10 km.

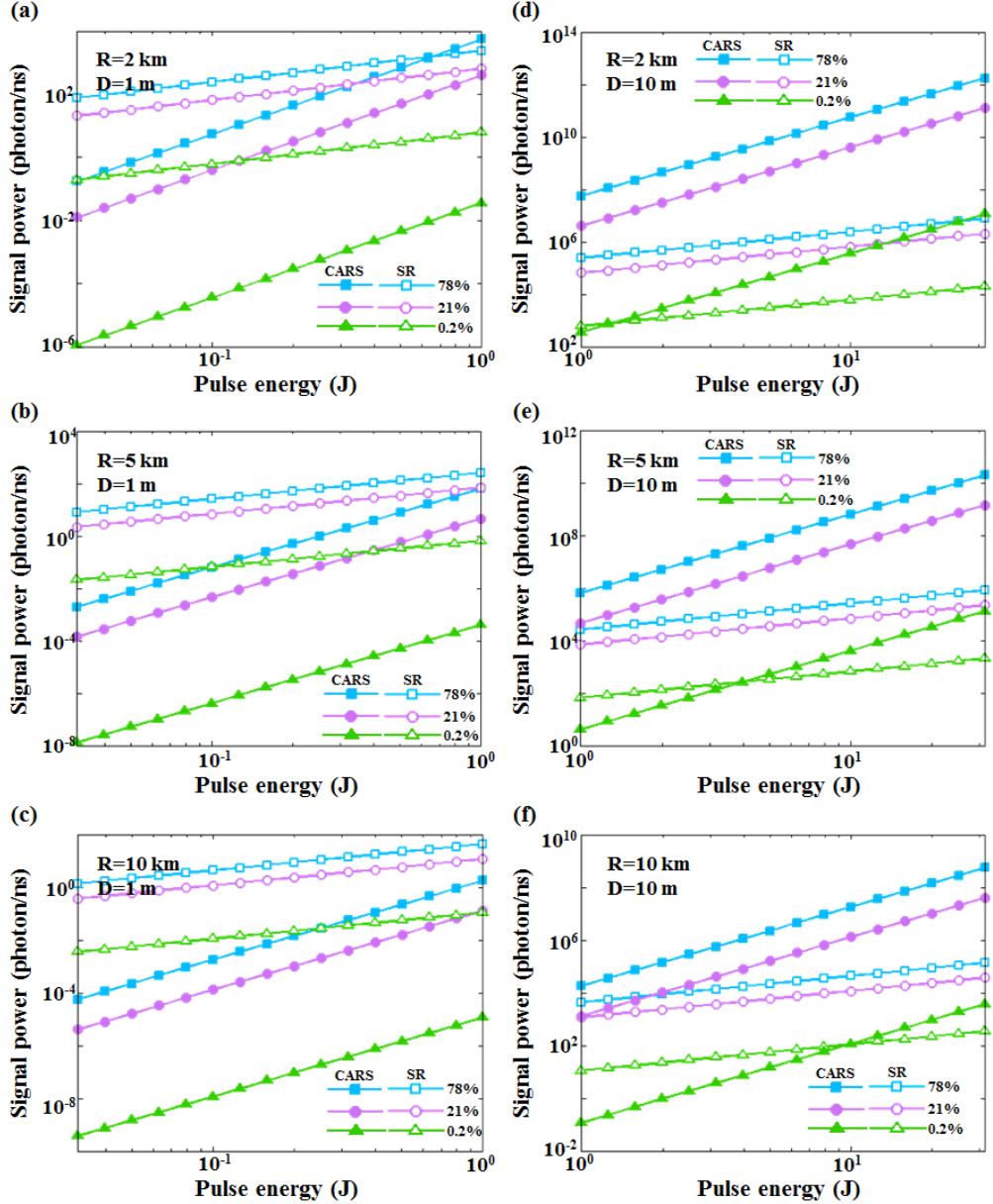

FIG. 4. Signal power of CARS lidar and spontaneous Raman (labelled by SR) lidar collected by a telescope with aperture diameter $D$=1 m and 10 m as a function of pulse energy for different percentages (78%, 21% and 0.2%) of the Raman medium at different altitude $R$=2 km, 5 km, and 10 km, respectively. SR: cyan open square (78%), magenta open circle (21%), green open triangle (0.2%); CARS: cyan full square (78%), magenta full circle (21%), green full triangle (0.2%). For CARS, the pulse energy means the total pulse energy of pump ad Stokes lasers. The wavelengths of both the spontaneous Raman signal and the CARS signal are assumed to be 500 nm.

## IV. CONCLUSION

We have proposed a scheme to enhance the backward CARS signal by a slow Stokes pulse generated by EIT effects, and we have investigated its potential application in a CARS lidar. We found that when the spatial length of the Stokes pulse is close to $0.13\lambda_a$, the backward CARS signal reaches its maximum. Normally, this requires the group velocity of the Stokes pulse to be as slow as $10^{-7} \sim 10^{-8} c$. Using $O_2$ or $N_2$ in the atmosphere as the EIT medium, our calculation seems to indicate that it is possible to achieve these conditions. We also found that under conventional Raman lidar parameters, the signal level of a CARS lidar is lower than that of a spontaneous Raman lidar. However, with higher pulse energy and a larger aperture telescope, CARS signal could surpass spontaneous Raman signal and a CARS lidar could become more sensitive than a spontaneous Raman lidar.

## APPENDIX:

Here we present a proof that the total momenta of the system are conserved no matter CARS photons are generated in the forward or backward direction.

The density of Lorentz force applied to the medium is $\boldsymbol{f} = \rho \boldsymbol{E} + \boldsymbol{j} \times \boldsymbol{B}$, where in dielectric medium, the charge density $\rho = -\nabla \cdot Re(\boldsymbol{P})$ and the current density $\boldsymbol{j} = \frac{\partial Re(\boldsymbol{P})}{\partial t}$. In CARS process, the polarization is

$$\begin{aligned} \boldsymbol{P}_a^{(3)} &= 3\varepsilon_0 \hat{\chi}^{(3)}(-\omega_a; \omega_p, \omega_p, -\omega_S) \cdot \boldsymbol{E}_p \boldsymbol{E}_p \boldsymbol{E}_S^* \\ \boldsymbol{P}_S^{(3)} &= 3\varepsilon_0 \hat{\chi}^{(3)}(-\omega_S; \omega_p, \omega_p, -\omega_a) \cdot \boldsymbol{E}_p \boldsymbol{E}_p \boldsymbol{E}_a^*. \\ \boldsymbol{P}_p^{(3)} &= 6\varepsilon_0 \hat{\chi}^{(3)}(-\omega_p; \omega_a, -\omega_p, \omega_S) \cdot \boldsymbol{E}_a \boldsymbol{E}_p^* \boldsymbol{E}_S \end{aligned} \quad (A1)$$

where $\hat{\chi}^{(3)}$ is a four dimension tensor, with full permutation symmetry [11]

$$\chi_{ijkl}^{(3)}(-\omega_a; \omega_p, \omega_p, -\omega_S) = \chi_{ljki}^{(3)}(-\omega_S; \omega_p, \omega_p, -\omega_a) = \chi_{jikl}^{(3)}(\omega_p; -\omega_a, \omega_p, -\omega_S) = \chi_{ijkl}^{(3)}. \quad (A2)$$

According to the reality of the fields [11], we have

$$\chi_{jikl}^{(3)}(\omega_p; -\omega_a, \omega_p, -\omega_S) = \chi_{jikl}^{(3)*}(-\omega_p; \omega_a, -\omega_p, \omega_S). \quad (A3)$$

Thus Eq. (A1) becomes

$$\begin{aligned} P_{a,i}^{(3)} &= 3\varepsilon_0 \sum_{jkl} \chi_{ijkl}^{(3)} E_{p,j} E_{p,k} E_{S,l}^* \exp(-i\omega_a t + i k_a z) \\ P_{S,l}^{(3)} &= 3\varepsilon_0 \sum_{ijk} \chi_{ijkl}^{(3)} E_{p,j} E_{p,k} [E_{a,i}^{FW*} \exp(-i\omega_S t + i k_S z) + E_{a,i}^{BW*} \exp(-i\omega_S t + i(2k_a + k_S)z)]. \\ P_{p,j}^{(3)} &= 6\varepsilon_0 \sum_{ikl} \chi_{ijkl}^{(3)*} E_{p,k}^* E_{S,l} [E_{a,i}^{FW} \exp(-i\omega_p t + i k_p z) + E_{a,i}^{BW} \exp(-i\omega_p t + i(k_p - 2k_a)z)] \end{aligned} \quad (A4)$$

where $E_{p,j}$, $E_{p,k}$, $E_{S,l}$ and $E_{a,i}$ are the envelop profiles of the pulse. $E_{a,i}^{BW}$ is complex [Eq. (8)], while $E_{p,j}$, $E_{p,k}$, $E_{S,l}$ and $E_{a,i}^{FW}$ are real [Eq.(2) and Eq. (7)].

$\boldsymbol{E}$ has only $x$ and $y$ component and varies along $z$ direction, thus $\nabla \cdot \boldsymbol{E} = 0$. According to Eq. (A1), $\rho = -\nabla \cdot \boldsymbol{P} = 0$. Then, the density of Lorentz force becomes

$$\boldsymbol{f} = \boldsymbol{j} \times Re(\boldsymbol{B}) = \frac{\partial Re(\boldsymbol{P})}{\partial t} \times [c\boldsymbol{n} \times Re(\boldsymbol{E})] = Re\left(\frac{\partial \boldsymbol{P}}{\partial t}\right) \cdot Re(\boldsymbol{E}) c\boldsymbol{n}, \quad (A5)$$

where $\boldsymbol{n}$ is the unit vector in the propagating direction. Since the envelop change slowly with time, the derivative with respect to time can be replaced by $i\omega$. The momentum transferred to the medium per unit cross-sectional area is $\Delta \boldsymbol{p} = \iint \boldsymbol{f} dt dz$. Since $\Delta \boldsymbol{p}$ is an integral over time, it vanishes unless $\boldsymbol{P}$ and $\boldsymbol{E}$ have the same frequency. Therefore,

$$\Delta \boldsymbol{p} = c\boldsymbol{n}_z \iint \left[ Re\left(-i\omega_a \boldsymbol{P}_a^{(3)}\right) \cdot Re(\boldsymbol{E}_a^{FW}) - Re\left(-i\omega_a \boldsymbol{P}_a^{(3)}\right) \cdot Re(\boldsymbol{E}_a^{BW}) \right. \\ \left. + Re\left(-i\omega_S \boldsymbol{P}_S^{(3)}\right) \cdot Re(\boldsymbol{E}_S) + Re\left(-i\omega_p \boldsymbol{P}_p^{(3)}\right) \cdot Re(\boldsymbol{E}_p) \right] dt dz \quad (A6)$$

where $\boldsymbol{n}_z$ is the unit vector in $z$ direction. Substituting Eq. (A4) into Eq. (A6), we have

$$\Delta \boldsymbol{p} = 3c\varepsilon_0 \boldsymbol{n}_z \\ \times \iint \sum_{ijkl} E_{p,j} E_{p,k} E_{S,l} \left\{ Im\left(\chi_{ijkl}^{(3)}\right) E_{a,i}^{FW} \left[\omega_a \cos^2(\omega_a \tau) + \omega_S \cos^2(\omega_S \tau) - 2\omega_p \cos^2(\omega_p \tau)\right] \right. \\ + \left|\chi_{ijkl}^{(3)}\right| \left|E_{a,i}^{BW}\right| \left[\omega_a \sin(\omega_a \tau - \varphi_{1,ijkl}) \cos(\omega_a \tau + 2k_a z - \varphi_{2,i}) \right. \\ \left. -\omega_S \sin(\omega_S \tau - 2k_a z - \varphi_{1,ijkl} + \varphi_{2,i}) \cos(\omega_S \tau) \right. \\ \left. \left. -2\omega_p \sin(\omega_p \tau + 2k_a z + \varphi_{1,ijkl} - \varphi_{2,i}) \cos(\omega_p \tau) \right] \right\} dt dz \quad (A6)$$

where $\tau = t - \frac{z}{c}$, $\varphi_1$ and $\varphi_2$ are defined by $\chi_{ijkl}^{(3)} = \left|\chi_{ijkl}^{(3)}\right| \exp(i\varphi_{1,ijkl})$ and $E_{a,i}^{BW} = \left|E_{a,i}^{BW}\right| \exp(i\varphi_{2,i})$. The integral in Eq. (A6) has two terms, the first one contains $E_{a,i}^{FW}$, while the second one contains $\left|E_{a,i}^{BW}\right|$. Since

$$\omega_a \cos^2(\omega_a \tau) + \omega_S \cos^2(\omega_S \tau) - 2\omega_p \cos^2(\omega_p \tau) = \frac{1}{2}[\omega_a \cos(2\omega_a \tau) + \omega_S \cos(2\omega_S \tau) - 2\omega_p \cos(2\omega_p \tau)], \quad (A7)$$

the first term related to forward CARS signal vanishes in integral with respect to $t$. Thus the momentum transfer to the medium is only caused by the generation of backward CARS photons. Since $\sin(\omega\tau)\cos(\omega\tau + \varphi) = \frac{1}{2}[-\sin(\varphi) + \sin(2\omega\tau + \varphi)]$, $\sin(2\omega\tau + \varphi)$ vanishes in integral with respect to $t$ as well, finally, Eq. (A6) becomes,

$$\Delta \boldsymbol{p} = 3\boldsymbol{n}_z c\varepsilon_0 \omega_a \iint \sum_{ijkl} \left|\chi_{ijkl}^{(3)}\right| E_{p,j} E_{p,k} E_{S,l} \left|E_{a,i}^{BW}\right| \sin(-2k_a z - \varphi_{1,ijkl} + \varphi_{2,i}) dt dz. \quad (A8)$$

The momentum flux density (momentum transferred through unit cross-sectional area per unit time) is $\varepsilon_0 c Re(\boldsymbol{E}_a^{BW}) \times Re(\boldsymbol{B}_a^{BW})$. Thus the total momentum of the backward propagating CARS field per unit cross-sectional area is

$$\boldsymbol{p}_a^{BW} = \lim_{z \to -\infty} \int \varepsilon_0 c Re(\boldsymbol{E}_a^{BW}) \times Re(\boldsymbol{B}_a^{BW}) dt \\ = -\boldsymbol{n}_z \varepsilon_0 c^2 \int \frac{\partial}{\partial z} \left[\int Re(\boldsymbol{E}_a^{BW}) \cdot Re(\boldsymbol{E}_a^{BW}) dt\right] dz \\ = \boldsymbol{n}_z c \iint Re\left(i\omega_a \boldsymbol{P}_a^{(3)}(z,t)\right) \cdot Re(\boldsymbol{E}_a^{BW}) dt dz \\ = -3\boldsymbol{n}_z c\varepsilon_0 \omega_a \iint \sum_{ijkl} \left|\chi_{ijkl}^{(3)}\right| E_{p,j} E_{p,k} E_{S,l}^* \sin(\omega_a t - k_a z - \varphi_{1,ijkl}) \left|E_{a,i}^{BW}\right| \cos(\omega_a t + k_a z - \varphi_{2,i}) dt dz \\ = -\frac{3}{2} \boldsymbol{n}_z c\varepsilon_0 \omega_a \iint \sum_{ijkl} \left|\chi_{ijkl}^{(3)}\right| E_{p,j} E_{p,k} E_{S,l}^* \left|E_{a,i}^{BW}\right| \sin(-2k_a z - \varphi_{1,ijkl} + \varphi_{2,i}) dt dz \quad (A9)$$

Here $\boldsymbol{E}_a^{BW} = -\frac{1}{2\varepsilon_0 c} \int_z^\infty \frac{\partial}{\partial t} \boldsymbol{P}_a^{(3)}\left(z', t + \frac{z-z'}{c}\right) dz'$ is used, and $\boldsymbol{P}_a^{(3)}$ is expressed in Eq. (A4). Comparing Eq. (A8) and (A9), we have

$$\Delta \boldsymbol{p} + 2\boldsymbol{p}_a^{BW} = 0. \quad (A10)$$

If a forward propagating CARS photon of momentum $\boldsymbol{p}_a$ is created, there will be two pump photons of momentum $\boldsymbol{p}_p$ annihilated and a Stokes photon of momentum $\boldsymbol{p}_S$ created, so that the momentum change of the photons is $\boldsymbol{p}_a - 2\boldsymbol{p}_p + \boldsymbol{p}_S = 0$, i.e., the momenta of photons are conserved. According to the discussion below Eq. (A7), there is no momentum transferred to the medium when a forward propagating CARS photon is created. Therefore, the total momenta of the system are apparently conserved.

On the other hand, if a backward propagating CARS photon with momentum $-\boldsymbol{p}_a$ is created, the momentum change of the photons is $-\boldsymbol{p}_a - 2\boldsymbol{p}_p + \boldsymbol{p}_S = -2\boldsymbol{p}_a$, twice of the momentum of the backward CARS photon. Thus in Eq. (A10), $2\boldsymbol{p}_a^{BW}$ is the total momentum change of photons per cross sectional area. Eq. (A10) means that it is completely compensated by the momentum transferred to the medium per cross sectional area. Therefore, the total momenta of the light-matter system are still conserved.

Furthermore, both $\Delta \boldsymbol{p}$ and $\boldsymbol{p}_a^{BW}$ contain the integral of $\sin(-2k_a z)$ with respect to $z$ [Eq. (A8-A9)]. This integral vanishes if the interaction length (range of $z$) is several times larger than the CARS wavelength. That is to say, the generation of backward CARS signal and the momentum transfer to the medium happen only when the interaction length is very small.